# MULTI-DISEASE DETECTION IN RETINAL IMAGING BASED ON ENSEMBLING HETEROGENEOUS DEEP LEARNING MODELS

*Dominik Müller[1], Iñaki Soto-Rey[1,2] and Frank Kramer[1]*

[1] IT-Infrastructure for Translational Medical Research, University of Augsburg, Germany
[2] Medical Data Integration Center, University Hospital Augsburg, Germany

**ABSTRACT**

Preventable or undiagnosed visual impairment and blindness affect billion of people worldwide. Automated multi-disease detection models offer great potential to address this problem via clinical decision support in diagnosis. In this work, we proposed an innovative multi-disease detection pipeline for retinal imaging which utilizes ensemble learning to combine the predictive capabilities of several heterogeneous deep convolutional neural network models. Our pipeline includes state-of-the-art strategies like transfer learning, class weighting, real-time image augmentation and Focal loss utilization. Furthermore, we integrated ensemble learning techniques like heterogeneous deep learning models, bagging via 5-fold cross-validation and stacked logistic regression models. Through internal and external evaluation, we were able to validate and demonstrate high accuracy and reliability of our pipeline, as well as the comparability with other state-of-the-art pipelines for retinal disease prediction.

*Index Terms—* Retinal Disease Detection, Ensemble Learning, Class Imbalance, Multi-label Image Classification, Deep Learning

## 1. INTRODUCTION

Even if the medical progress in the last 30 years made it possible to successfully treat the majority of diseases causing visual impairment, growing and aging populations lead to an increasing challenge in retinal disease diagnosis [1]. The World Health Organization (WHO) estimates the prevalence of blindness and visual impairment to 2.2 billion people worldwide, of whom at least 1 billion affections could have been prevented or is yet to be addressed [2]. Early detection and correct diagnosis are essential to forestall disease course and prevent blindness.

The use of clinical decision support (CDS) systems for diagnosis has been increasing over the past decade [3]. Recently, modern deep learning models allow automated and reliable classification of medical images with remarkable accuracy comparable to physicians [4]. Nevertheless, these models often lack capabilities to detect rare pathologies such as central retinal artery occlusion or anterior ischemic optic neuropathy [5], [6].

In this study we push towards creating a highly accurate and reliable multi-disease detection pipeline based on ensemble, transfer and deep learning techniques. Furthermore, we utilize the new Retinal Fundus Multi-Disease Image Dataset (RFMiD) containing various rare and challenging conditions to demonstrate our detection capabilities for uncommon diseases.

## 2. METHODS

The implemented medical image classification pipeline can be summarized in the following core steps and is illustrated in Fig. 1:
- Stratified multi-label 5-fold cross-validation
- Class weighted Focal loss and up-sampling
- Extensive real-time image augmentation
- Multiple deep learning model architectures
- Ensemble learning strategies: bagging and stacking
- Individual training for multi-disease labels and disease risk detection utilizing transfer learning on ImageNet
- Stacked binary logistic regression models for distinct classification

### 2.1. Retinal Imaging Dataset

The RFMiD dataset consists of 3200 retinal images for which 1920 images were used as training dataset [7]. The fundus images were captured by three different fundus cameras having a resolution of 4288x2848 (277 images), 2048x1536 (150 images) and 2144x1424 (1493 images), respectively.

**Tab. 1.** Annotation frequency for each class in the dataset.

| Disease | Samples | Disease | Samples | Disease | Samples |
|---|---|---|---|---|---|
| D. Risk | 1519 | DR | 376 | ARMD | 100 |
| MH | 317 | DN | 138 | MYA | 101 |
| BRVO | 73 | TSLN | 186 | ERM | 14 |
| LS | 47 | MS | 15 | CSR | 37 |
| ODC | 282 | CRVO | 28 | TV | 6 |
| AH | 16 | ODP | 65 | ST | 5 |
| AION | 17 | PT | 11 | RT | 14 |
| RS | 43 | CRS | 32 | EDN | 15 |
| RPEC | 22 | MHL | 11 | RP | 6 |
| OTHER | 34 | | | | |

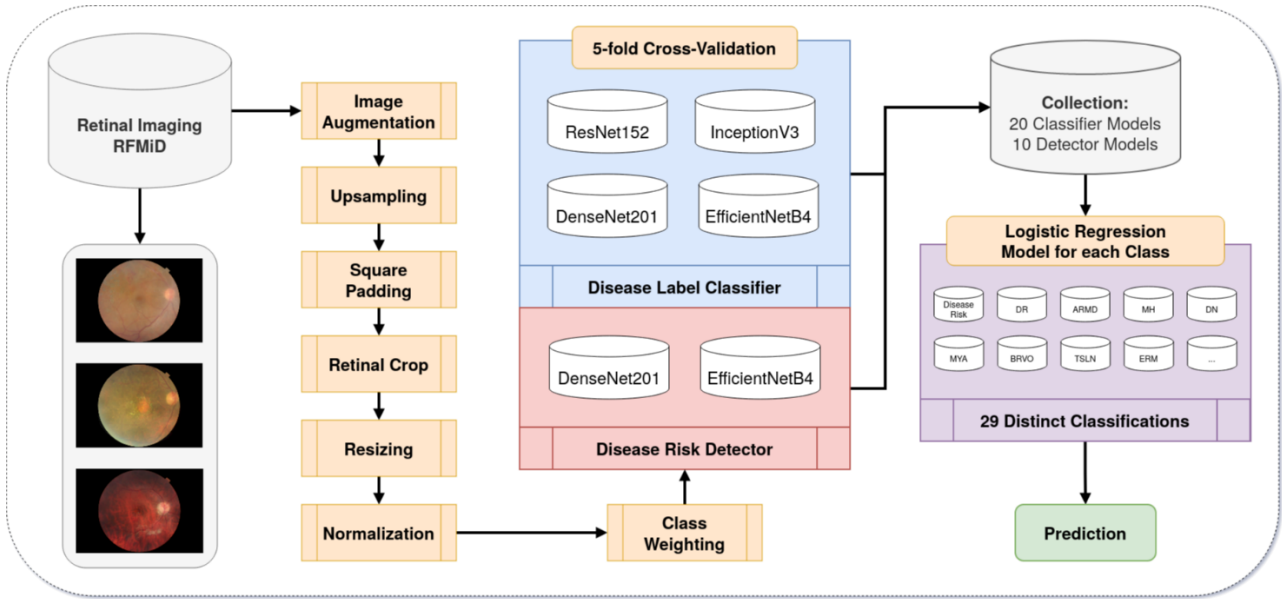

**Fig. 1**. Flowchart diagram of the implemented medical image analysis pipeline for multi-disease detection in retinal imaging. The workflow is starting with the retinal imaging dataset (RFMiD) and ends with computed predictions for novel images.

The images were annotated with 46 conditions, including various rare and challenging diseases, through adjudicated consensus of two senior retinal experts. These 46 conditions are represented by the following classes, which are also listed in Tab. 1: An overall normal/abnormal class, 27 specific condition classes and 1 'OTHER' class consisting of the remaining extremely rare conditions. Besides the training dataset, the organizers of the RIADD challenge hold 1280 images back for external validation and testing datasets to ensure robust evaluation [7], [8].

### 2.2. Preprocessing and Image Augmentation

In order to simplify the pattern finding process of the deep learning model, as well as to increase data variability, we applied several preprocessing methods.

We utilized extensive image augmentation for up-sampling to balance class distribution and real-time augmentation during training to obtain novel and unique images in each epoch. The augmentation techniques consisted of rotation, flipping, and altering in brightness, saturation, contrast and hue. Through the up-sampling, it was ensured that each label occurred at least 100 times in the dataset which increased the total number of training images from 1920 to 3354.

Afterwards, all images were square padded in order to avoid aspect ratio loss during posterior resizing. The retinal images were also cropped to ensure that the fundus is center located in the image. The cropping was performed individually for each microscope resolution and resulted in the following image shapes: 1424x1424, 1536x1536 and 3464x3464 pixels. The images were then resized to model input sizes according to the neural network architecture, which was 380x380 for EfficientNetB4, 299x299 for InceptionV3 and 244x244 for all remaining architectures [9]–[12].

Before feeding the image to the deep convolutional neural network, we applied value intensity normalization as last preprocessing step. The intensities were zero-centered via the Z-Score normalization approach based on the mean and standard deviation computed on the ImageNet dataset [13].

### 2.3. Deep Learning Models

The state-of-the-art for medical image classification are the unmatched deep convolutional neural network models [4], [14]. Nevertheless, the hyper parameter configuration and architecture selection are highly dependent on the required computer vision task, as well as the key difference between pipelines [4], [15]. Thus, our pipeline combines two different types of image classification models: The disease risk detector for binary classifying normal/abnormal images and the disease label classifier for multi-label annotation of abnormal images.

Both model types were pretrained on the ImageNet dataset [13]. For the fitting process, we applied a transfer learning training, with frozen architecture layers except for the classification head, and a fine-tuning strategy with unfrozen layers. Whereas the transfer learning fitting was performed for 10 epochs using the Adam optimization with an initial learning rate of 1-E04, the fine-tuning had a maximal training time of 290 epochs and using a dynamic learning rate for the Adam optimization starting from 1-E05 to a maximum decrease to 1-E07 (decreasing factor of 0.1 after 8 epochs without improvement on the monitored validation loss) [16]. Furthermore, an early stopping and model checkpoint technique was utilized for the fine-tuning process, stopping after 20 epochs without improvement (after epoch 60) and saving the best model measured according to the validation loss. Instead of defining an epoch as a cycle through the full training dataset, we establish an epoch to have 250 iterations. This allowed to increase the number of seen batches and, thus, to increase the information given to the model during the fitting process of an epoch. As training loss function, we utilized the weighted Focal loss from *Lin et al.* [17].

$$\text{FL}(p_t) = -\alpha_t(1-p_t)^\gamma \log(p_t) \quad (1)$$

In the above formula, $p_t$ is the probability for the correct ground truth class $t$, $\gamma$ a tunable focusing parameter (which we set to 2.0) and $\alpha_t$ the associated weight for class $t$.

*2.3.1 Disease Risk Detector*

The disease risk detector was established as a binary classifier of the disease risk class for general categorizing between normal and abnormal retinal images. Thus, this model type was trained using only the disease risk class and ignoring all multi-label annotations. Rather than using a single model architecture, we trained multiple models based on the DenseNet201 and EfficientNetB4 architecture [9], [10]. For class weight computation, we divided the number of samples by the multiplication of the number of classes (2 for a binary classification) with the number of class occurrences in the dataset.

*2.3.2 Disease Label Classifier*

In contrast, the disease label classifier was established as multi-label classifier of all 28 remaining classes (excluding disease risk) and was trained on the one hot encoded array of the disease labels. Furthermore, we utilized four different architectures for this model type: ResNet152, InceptionV3, DenseNet201 and EfficientNetB4 [9]–[12]. Identical to class weight computation of the disease risk detector, we computed the weights individually as binary classification for each class. Even if this classifier is provided with all classes, the binary weights balance the decision for each label individually.

## 2.4. Ensemble Learning Strategy

*2.4.1 Bagging*

Next to the utilization of multiple architecture, we also applied a 5-fold cross-validation based as a bagging approach for ensemble learning. Our aim was to create a large variety of models which were trained on different subsets of the training data. This approach not only allowed a more efficient usage of the available training data, but also increased the reliability of a prediction. This strategy resulted in an ensemble of 10 disease risk detector models (2 architectures with each 5 folds) and 20 disease label classifier models (4 architectures with each 5 folds).

*2.4.2 Stacking*

For combining the predictions of our, in total, 30 models, we integrated a stacking setup. On top of all deep convolutional neural networks, we applied a binary logistic regression algorithm for each class, individually. Thus, the predictions of all models were utilized as input for computing the classification of a single class. This approach allowed combining the information of all other class predictions to derive an inference for one single class. Overall, this strategy resulted in 29 distinct logistic regression models (1 for disease risk and 28 for each disease-label including the 'other' class). The individual predicted class probabilities are then concatenated to the final prediction.

The logistic regression models were also trained with the same 5-fold cross-validation sampling on a heavily augmented version of the training dataset to avoid overfitting as well as avoiding training the logistic regression models on already seen images from the neural network models. As logistic regression solver, we utilized the large-scale bound-constrained optimization (short: 'LBFGS') from *Zhu et al.* [18].

## 3. RESULTS AND DISCUSSION

The sequential training of a complete cross-validation for one architecture on a single NVIDIA TITAN RTX GPU took around 13.5 hours with 63 epochs on average for each deep convolutional neural network model. Logistic Regression training required less than 30 minutes for all class models combined. No signs of overfitting were observed for the disease label classifiers through validation monitoring, as it can be seen in Fig. 2. However, the disease risk detectors showed a strong trend to overfit after the transfer learning phase. Through our strategy to use the model with the best

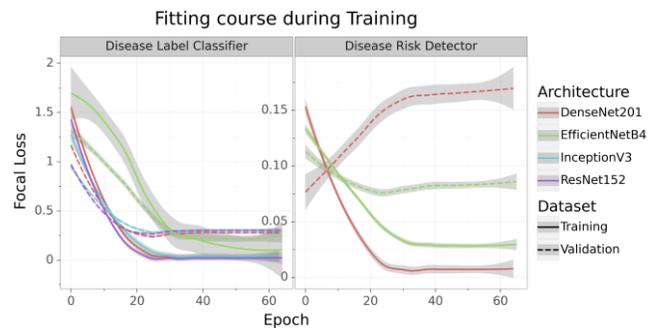

**Fig. 2.** Loss course during the training process for training and validation data. The lines were computed via locally estimated scatterplot smoothing and represent the average loss across all folds. The gray areas around the lines represent the confidence intervals.

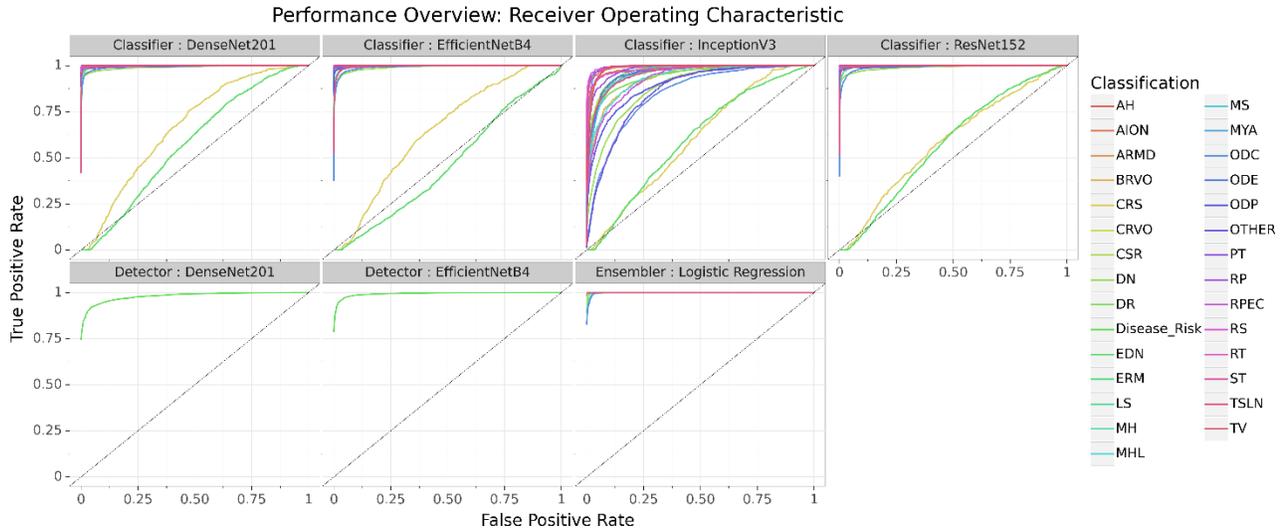

**Fig. 3.** Receiver operating characteristic (ROC) curves for each model type applied in our pipeline. The ROC curves showing the individual model performance measured by the true positive and false positive rate. The cross-validation models were macro-averaged for each model type to reduce illustration complexity.

validation loss, it was still possible to obtain a powerful model for detection.

### 3.1. Internal Performance Evaluation

For estimating the performance of our pipeline, we utilized the validation subsets of the 5-fold cross-validation models from the heavily augmented version of our dataset. This approach allowed to obtain testing samples which were never seen in the training process for reliable performance evaluation. For the complex multi-label evaluation, we computed the popular area under the receiver operating characteristic (AUROC) curve, as well as the mean average precision (mAP). Both scores were macro-averaged over classes and cross-validation folds to reduce complexity.

Our multi-disease detection pipeline revealed a strong and robust classification performance with the capability to also detect rare conditions accurately in retinal images. Whereas the disease label classifier models separately only achieved an AUROC of around 0.97 and a mAP of 0.93, the disease risk detectors demonstrated to have a really strong predictive power of 0.98 up to 0.99 AUROC and mAP. However, for the classifiers the InceptionV3 architecture indicated to have the worst performance compared to the other architectures with only 0.93 AUROC and 0.66 mAP. The associated receiver operating characteristics of the models are illustrated in Fig. 3.

Training a strong multi-label classifier is in general a complex task, however, the extreme class imbalance between the conditions revealed a hard challenge for building a reliable model [19], [20]. Our applied up-sampling and class weighting technique demonstrated to have a critical boost on the predictive capabilities of the classifier models. Nearly all labels were able to be accurately detected, including the 'OTHER' class consisting of various extremely rare conditions. Nevertheless, the two classes 'EDN' and 'CRS' were the most challenging conditions for all classifier models. Both classes belong to very rare conditions, combined with less than 1.2% occurrence in the original and 2.5% occurrence in the up-sampled dataset. Still, our stacked logistic regression algorithm was able to balance this issue and infer the correct 'EDN' and 'CRS' classifications through context. Overall, our applied ensemble learning strategies resulted in a significant performance improvement compared to the individual deep convolutional neural network models. More details on the internal performance evaluation are listed in Tab. 2.

### 3.2. External Evaluation through the RIADD Challenge

Furthermore, we participated at the RIADD challenge which was organized by the authors of the RFMiD dataset [7], [8]. The challenge participation allowed not only an independent evaluation of the predictive power of our pipeline on an unseen and unpublished testing set, but also the comparison with the currently best retinal disease classifiers in the world.

**Tab. 2**. Achieved results of the internal performance evaluation showing the average AUROC and mAP score for each model utilized in our pipeline. The scores were macro-averaged across all cross-validation folds and classes.

| Model Type | Architecture | AUROC | mAP |
|---|---|---|---|
| **Classifier** | DenseNet201 | 0.973 | 0.931 |
| **Classifier** | EfficientNetB4 | 0.969 | 0.929 |
| **Classifier** | ResNet151 | 0.970 | 0.930 |
| **Classifier** | InceptionV3 | 0.932 | 0.663 |
| **Detector** | DenseNet201 | 0.980 | 0.997 |
| **Detector** | EfficientNetB4 | 0.993 | 0.999 |
| **Ensembler** | Logistic Regression | 0.999 | 0.999 |

In our participation, we were able to reach rank 19 from a total of 59 teams in the first evaluation phase and rank 8 in the final phase. In the independent evaluation from the challenge organizers, we achieved an AUROC of 0.95 for the disease risk classification. For multi-label scoring, they computed the average between the macro-averaged AUROC and the mAP, for which we reached the score 0.70. The top performing ranks shared only a marginal scoring difference which is why we had only a final score difference of 0.05 to the first ranked team. Furthermore, the participation results demonstrated that ensemble learning based classification for deep convolutional neural network models is compatible or even superior to other approaches in the scientific field such as focusing on a single large architecture.

### 3.3. Experiments and Improvements

Additionally, we experimented with using weighted cross-entropy loss for training our both model types. This resulted in inferior models for disease label classification, however, the cross-entropy loss fitted disease risk detector models showed less overfitting with equal performance. Further experimentation with loss functions for the disease risk detector models could provide the solution to avoid overfitting.

An important point for the RIADD challenge participation would be the utilization of more training data, especially for the difficult 'CRS' and 'EDN' classes. According to the challenge rules, other public available datasets like Kaggle DR, IDRiD, Messidor or APTOS are allowed to be used as additional training data [8]. Our pipeline, which was trained exclusively on the RFMiD dataset, could be further improved with more retinal images of very rare conditions. Besides the training data, more improvement points for further research in retinal disease detection would be the inclusion of image cropping strategies to reduce information loss through resolution resizing, the usage of more architectures (especially with different input resolutions) to increase the model ensemble, and the utilization of specific retinal filters or retinal vessel segmentation as additional information to utilize for the predictions.

### 4. CONCLUSIONS

In this study, we introduced a powerful multi-disease detection pipeline for retinal imaging which exploits ensemble learning techniques to combine the predictions of various deep convolutional neural network models. Next to state-of-the-art strategies, such as transfer learning, class weighting, extensive real-time image augmentation and Focal loss utilization, we applied 5-fold cross-validation as bagging technique and used multiple convolutional neural network architectures to create an ensemble of models. With a stacking approach of class-wise distinct logistic regression models, we combined the knowledge of all neural network models to compute highly accurate and reliable retinal condition predictions. Next to an internal performance evaluation, we also proved the precision and comparability of our pipeline through the participation at the RIADD challenge.

### APENDIX

In order to ensure full reproducibility and to create a base for further research, the complete code of this study, including extensive documentation, is available in the following public Git repository:
https://github.com/frankkramer-lab/riadd.aucmedi
Furthermore, the trained models, evaluation results and metadata are available in the following public Zenodo repository:
https://doi.org/10.5281/zenodo.4573990

### ACKNOWLEDGMENTS

We want to thank Dennis Klonnek, Edmund Müller and Johann Frei for their useful comments and support.

### COMPLIANCE WITH ETHICAL STANDARDS

This research study was conducted retrospectively using human subject data made available in open access by *Pachade et al.* [7], [8]. Ethical approval was not required as confirmed by the license attached with the open access data.

### CONFLICT OF INTEREST

None declared.

### FUNDING

This work is a part of the DIFUTURE project funded by the German Ministry of Education and Research (Bundesministerium für Bildung und Forschung, BMBF) grant FKZ01ZZ1804E.